\documentclass{aa}  
\usepackage[varg]{txfonts}
\usepackage{graphicx}
\usepackage{txfonts}
\usepackage{lipsum}
\usepackage{subcaption}         
\usepackage{lscape}             
\usepackage{placeins}    
\usepackage{amsfonts}
\usepackage{amssymb}
\usepackage{amsmath}
\usepackage{natbib}
\usepackage{dcolumn}
\usepackage{bm} 
\usepackage{color,xcolor}
\usepackage{multirow}
\usepackage{caption}
\usepackage{hyperref}
\usepackage[normalem]{ulem}
\usepackage[export]{adjustbox}
\usepackage{float}
\usepackage{scrextend}
\usepackage{textgreek}
\usepackage{cuted}

\renewcommand{\linenumbers}{} 

\begin{document}

\title{Role of the $\delta$ Meson in the Equation of State\\and Direct Urca Cooling of Neutron Stars}

 \author{L. Scurto\inst{1}\fnmsep\thanks{lscurto@student.uc.pt}
        \and H. Pais\inst{1}\fnmsep\thanks{hpais@uc.pt}
        \and M. Antonelli\inst{2}\fnmsep\thanks{antonelli@lpccaen.in2p3.fr}
        \and F. Gulminelli\inst{2}\fnmsep\thanks{gulminelli@lpccaen.in2p3.fr}
        }

   \institute{CFisUC, Department of Physics, University of Coimbra, 3004-516 Coimbra, Portugal\\
            \and Normandie Univ., ENSICAEN, UNICAEN, CNRS/IN2P3, LPC Caen, F-14000 Caen, France\\ }

\date{}

\abstract 

{The direct Urca (dUrca) process is a key mechanism driving rapid neutrino cooling in neutron stars, with its baryon density activation threshold determined by the microscopic model for nuclear matter. Understanding how nuclear interactions shape the dUrca threshold is essential for interpreting neutron star thermal evolution, particularly in light of recent studies on exceptionally cold objects.}  
{We investigate the impact of incorporating the scalar isovector $\delta$ meson into the neutron star equation of state, which alters the internal proton fraction and consequently affects the dUrca cooling threshold. Since proton superfluidity is known to suppress dUrca rates, we also examine the interplay between the nuclear interaction mediated by the $\delta$ meson and the $^1S_0$ proton pairing gap.}  
{We perform a Bayesian analysis using models built within a relativistic mean-field approximation, incorporating constraints from astrophysical observations, nuclear experiments, and known results of \textit{ab initio} calculations of pure neutron matter. We then impose a constraint on the dUrca threshold based on studies of fast-cooling neutron stars.}  
{The inclusion of $\delta$ meson expands the range of possible internal compositions, directly influencing the stellar mass required for the central density to reach the dUrca threshold. Furthermore, we observe that the observation of relatively young and cold neutron stars provides insights into $^1S_0$ proton superfluidity in the core of neutron stars.
}
{}

\maketitle

\section{Introduction}

The advent of multi-messenger astronomy, particularly the detection of gravitational waves, has provided new tools to investigate the equation of state (EoS) of neutron stars (NSs) and renewed interest in understanding their internal composition.  
The observation of GW170817 by \cite{Abbott_2017,Abbott_2019}, combined with mass and radius inferences of PSR J0030+0451 \citep{Riley_2019,Miller_2019}, PSR J0740+6620 \citep{Riley_2021,Miller_2021}, and PSR J0437+4715 \citep{Choudhury_2024,Choudhury_zenodo} by the NICER experiment, has placed constraints on the high-density regime of the EoS. These constraints can be complemented by information from the low-density regime provided by nuclear experiments and \textit{ab initio} calculations , see \citet{PREX_2021,CREX_2022},~\citet{Huth_2021,Hebeler_2013}.
These constraints have been extensively used in the literature to largely explore the EoS parameter space, employing a variety of approaches. Agnostic methods include piecewise polytropes \citep{Read_2008}, spectral parametrizations \citep{Lindblom_2010,Lindblom_2012,Lindblom_2013}, Gaussian process-based sampling \citep{Landry_2018,Essick_2019,Landry_2020}, and neural networks \citep{Han_2023}. Other approaches involve non-relativistic  \citep{Margueron_2018_prc,Mondal_2023,Montefusco_2025} and relativistic ~\citep{Providencia_2019,Char_2023,Malik_2024,Scurto_2024} mean-field  calculations.

A recent analysis by \cite{Marino_2024}, based on the analysis of data from XMM-Newton and Chandra suggests that the direct Urca (dUrca) process may be necessary to explain the observed temperatures of three exceptionally cold and relatively young NSs. The presence of such fast cooling objects had already been suggested by the study of isolated neutron stars in \citet{Potekhin2015review,Potekhin_2018}, as well as of transiently accreting neutron stars \citep{Potekhin_2023}. The existence of these objects may place non-trivial constraints on the neutron star EoS, as enhanced cooling mechanisms -- such as neutrino emission through dUrca processes -- are required to fit theoretical models to observational data. 
While such enhanced cooling might alternatively be explained with the presence of exotic matter such as hyperons or quark matter \citep{Prakash_1992,Prakash_1996,Fortin_2021}, the possibility of it being due to nucleonic direct Urca process needs to be studied in detail.
In fact,
proton superfluidity in the core is known to suppress fast dUrca processes, see \citet{Wei_2019,Sedrakian2019EPJA}, if the stellar temperature is below the pairing gap, which in turn depends on the matter composition, and therefore on the EoS. 
To investigate the complex interplay between EoS and composition, we introduce hypothetical constraints on the mass threshold for the onset of the dUrca process and analyze how these constraints affect the posterior distributions of astrophysical observables, EoS and pairing properties of nuclear matter. 

To this end, we follow 
the approach of~\cite{Wei_2020,Das_2024}
concerning pairing, and assume a simplified model for the $^1S_0$ proton pairing gap, essentially a rescaled version of the BCS pairing gap computed in ~\cite{Lombardo_2000}.

To account for the uncertainties in the composition, we build a large family of stable and causal nuclear EoSs
within the RMF approach, where the nuclear interaction is mediated by the exchange of virtual mesons.
In most applications of this formalism, see \citet{Serot:1984ey,Providencia_2019,Pais_2015,Char_2023,Scurto_2024,Malik_2024}, three mesons are typically considered: the scalar-isoscalar meson $\sigma$, the vector-isoscalar meson $\omega$, and the vector-isovector meson $\rho$. However, in this standard implementation, the absence of isospin dependence in the mass term leads to identical effective Dirac masses for protons and neutrons, which is not expected from microscopic calculations, see \citet{VanDalen_2005,Wang_2023}. 
To address this limitation, several studies  have introduced a fourth meson, the scalar isovector meson $\delta$, see \citet{Liu_2002,Gaitanos_2004,Menezes_2004,Pais_2009,Roca-Maza_2011}. The $\delta$ meson also plays a strong role in the symmetry energy and its derivatives \cite{Liu_2002,Gaitanos_2004}, and this has an impact in the proton fraction of the system.

If the existence of a mass splitting is firmly established, its actual value and density dependence is largely unconstrained, see for instance \cite{Margueron_2018} for a discussion.
The impact of the $\delta$ meson on the EoS has been recently investigated in \cite{Santos_2024}, where the authors selected three models without the $\delta$ meson from previous works and extended them by including this additional meson. They found that this interaction channel has a strong effect on the macroscopic properties of the stars, namely the radius and tidal deformability of low- and intermediate-mass stars, and also on their proton content in the core, which is the key variable for the possibility of nucleonic dUrca. 

Motivated by these results, we conduct a comprehensive Bayesian analysis on two sets of RMF models (with and without the $\delta$ meson) and largely vary its unknown effective coupling to the nucleons, in order to understand the role of this isovector scalar field on the properties of the stars.
In particular, we show that the extra inclusion of the $\delta$ meson enhances the flexibility of the models   allowing for a broader range of proton fractions at high densities, with respect to previous Bayesian studies (see for instance the one in \cite{Malik_2022}).

The structure of the paper is as follows. In the next section, we present the relativistic mean-field (RMF) formalism adopted in this study. Section~\ref{Sec_3} outlines the framework used for the Bayesian analysis. In Section~\ref{Sec_4}, we discuss the results of the comparison between the models with and without the $\delta$ meson. 
Section~\ref{Sec_5} introduces the constraints on the dUrca threshold and the proton pairing gap and examines their effects. Finally, in Section~\ref{Sec_6}, we summarize our results.
A more detailed of the RMF used in the study is shown in App.\ref{App_RMF}.

\section{Construction of the Equation of State}
\label{Sec_2}

To set up our family of EoSs, we work within the RMF approximation, where the interaction between nucleons is mediated by virtual mesons. Specifically, we compare the results obtained from two sets of EoS models, each based on a different structure of the Lagrangian density that describes the system. 
The first set, referred to as Set A, includes three mesonic fields mediating the nucleon interactions: the scalar isoscalar meson $\sigma$, the vector isoscalar meson $\omega$, and the vector isovector meson $\rho$. The second set, referred to as Set B, extends this framework by incorporating an additional meson, the scalar isovector meson $\delta$. 
In the following sections, we will discuss the results obtained from these two sets and explore the effects of the $\delta$ meson on the EoS and astrophysical predictions.

Our RMF procedure is based on the standard phenomenological Lagrangian density form :

\begin{equation}
\mathcal{L}=\mathcal{L}_N+\mathcal{L}_L+\mathcal{L}_\sigma+\mathcal{L}_\omega+\mathcal{L}_\rho +\mathcal{L}_\delta \, ,
\label{Eq_1}
\end{equation}

where $\mathcal{L}_L$ is the leptonic contribution and $\mathcal{L}_{\sigma,\omega,\rho,\delta}$ correspond to the mesonic fields, see App.~\ref{App_RMF} for details.
Here we focus on the first term, which describes the nucleons:

\begin{equation}
        \mathcal{L}_N=\bar{\psi}
        \left[
        \gamma_\mu  i \partial^\mu-g_\omega V^\mu-\frac{g_\rho}{2}\boldsymbol{\tau}\cdot\mathbf{b}^\mu-M^*
        \right]\psi \, ,
\end{equation}

where the Dirac field $\psi=(\psi_n,\psi_p)^T$ carries an isospin index ($\boldsymbol{\tau}$ are Pauli matrices acting in isospin space) and 

\begin{equation}
\label{Meffective}
    M^* = M-g_\sigma\phi + \frac{g_\delta}{2}\boldsymbol{\tau}\cdot\boldsymbol{\delta}
\end{equation}

is the effective nucleon mass, with $M=M_p=M_n=938.9\,$MeV the bare nucleon mass.  
 The main effect of the isoscalar vector field, $\boldsymbol{\delta}=(0,0,\delta_3)$, is to introduce an isospin dependence in the effective mass, $M^*=\text{diag}(M_n,M_p)$ with 

\begin{equation}
\label{Meffective3}
    M^*_n = M-g_\sigma\phi + \frac{g_\delta}{2}\delta_3 \, ,
    \qquad 
    M^*_p = M-g_\sigma\phi - \frac{g_\delta}{2}\delta_3    \, ,
\end{equation}

for the neutrons and protons, respectively, see for instance ~\cite{Liu_2002,Menezes_2004,Pais_2009,Roca-Maza_2011}.
The case without the additional meson can be recovered by switching off the coupling between the $\delta$ meson and the nucleons by setting~$g_\delta=0$: this automatically implies that $\delta_3=0$, see Eq.~\eqref{EL_delta}.
 
Furthermore, we consider the nucleon-meson couplings as functions of the total baryonic density by adopting the density dependence previously used in~\cite{Malik_2022}: 
\begin{equation}
\label{gabc}
    g_i = a_i + b_i \, e^{-c_i \, x}  
    \qquad \qquad \quad (i=\sigma,\rho,\omega,\delta) 
\end{equation}

where $x=n/n_{sat}$ is the baryon number density $n$ measured in units of  $n_{sat}$, the saturation density of symmetric matter. 
The parameters $a_i$, $b_i$, and $c_i$ are free and will be varied to sample various nuclear models: their specific value defines each model within the two sets, A and B.
Hence, each model is uniquely specified by 9 parameters for Set A (without the $\delta$ meson), and 12 parameters for~Set~B (with the $\delta$ meson, that induces the effective mass splitting). 

We will demonstrate that incorporating the $\delta$ meson allows the density dependence in \eqref{gabc} to reproduce nearly the entire range of results obtained in \cite{Char_2023,Scurto_2024} who used the more general GDFM formulation of \cite{Gogelein_2008}. \footnote{The difference between the two approaches lies in the additional $x^3$ term present in the GDFM expression for the $g_i$ parameters. Here, the choice of different density-dependent parameters $g_i$ compared to that of \cite{Char_2023,Scurto_2024} is motivated by the need to avoid the irregular behaviour of the symmetry energy induced by the cubic term, as observed in~\cite{Scurto_2024}.}

\section{Bayesian Setup}\label{Sec_3}

We now present the methodology used for our Bayesian analysis, which closely follows the approach detailed in~\cite{Scurto_2024}.

\begin{table}[]
    \centering
    \begin{tabular}{c|cc|cc}
        \hline
        \hline
         & \multicolumn{2}{c}{Set A} &   \multicolumn{2}{c}{Set B} \\
         & Min & Max & Min & Max \\
         \hline
          $n_{sat}\,$(fm$^{-3})$ & 0.140 & 0.170 & 0.140 & 0.170 \\
          $E_{sat}\,$(MeV) & -17 & -14 & -17 & -14\\
          $J_{sym}\,$(MeV) & 28  & 36  & 28  & 36\\
          $L_{sym}\,$(MeV) & 0   & 120 & 0   & 120\\
          $b_\sigma$ & 1.8 & 2.4 & 1.8 & 2.4 \\
          $c_\sigma$ & 2.0 & 3.0 & 2.0 & 3.0 \\
          $b_\omega$ & 2.0 & 2.4 & 2.0 & 2.4\\
          $c_\omega$ & 2.0 & 3.0 & 2.0 & 3.0\\
          $b_\rho$   & 2.0 & 6.0 & 2.0 & 6.0\\
          $c_\rho$   & /   & /   & 0.0 & 2.0\\
          $b_\delta$ & /   & /   & 0.0 & 10.0\\
          $c_\delta$ & /   & /   & 0.0 & 2.0\\
          \hline
          \hline
    \end{tabular}
    \caption{The minimum and maximum values of the uniform prior distributions of the nuclear matter properties and couplings for Sets A and B.}
   \label{Tab_generation}
\end{table}

\subsection{Construction of the Prior}

We begin by outlining the procedure used to construct the prior distribution. As discussed above, our goal is to compare results obtained from two prior sets: Set A, where the interaction mediated by the $\delta$ meson is switched off, and Set B, where it is accounted for. 

Following \cite{Scurto_2024}, instead of directly sampling all coupling parameters that appear in $\mathcal{L}$, we sample a combination of nuclear matter parameters (NMPs) and coupling parameters using uniform distributions, as detailed in Tab.~\ref{Tab_generation}. The remaining coupling parameters are then computed using analytical relations which can be found in App.\ref{App_NMP}. The chosen parameters exhibit minimal correlations, and the sampling ranges are set to explore a broad parameter space while ensuring physically meaningful models.

The sampled models are then filtered based on the following criteria: each equation of state must have a monotonically increasing energy density and pressure for both neutron and beta-equilibrated matter, as well as showing only the expected minimum in the energy per baryon of symmetric matter at saturation, and must yield a minimum $\chi^2$ in the nuclear mass fit used for crust calculations,  see ~\cite{Scurto_2024} for details. After filtering, we obtain two prior distributions, each containing $10^6$ models.

\subsection{Posterior Distribution}

With the prior distributions established, we derive the posterior distributions by incorporating constraints from astrophysical observations, nuclear experiments, and ab initio calculations. 
The posterior probability for a given EoS is:
\begin{equation}
        P({\mathbf X}|{\mathbf c}) \propto \prod_k P(c_k|{\mathbf X}) \, ,
\end{equation}
where $\bf{X}$ represents the 9 (12) parameters defining the EoS (i.e., the ones listed in Tab.~\ref{Tab_generation}), and the label $k$ runs over the set of constraints~$\mathbf c$. The probability distributions for observables $Y({\bf X})$ are obtained through marginalization over the parameter space ($N=9$ for Set A, $N=12$ for Set B):
\begin{equation}
        P(Y|{\mathbf c})=
        \prod_{j=1}^N\int_{X_j^{min}}^{X_j^{max}}
        \!\!\!\! dX_j \, 
	P({\mathbf X}|{\mathbf c}) \, 
    \delta \left( Y-Y({\mathbf X}) \right).
\end{equation}
In the following, we briefly recall the constraints used.

\subsubsection{Astrophysical Constraints}

The astrophysical constraints are the same as in~\cite{Scurto_2024,davis2024,Montefusco_2025}, including:
\begin{itemize}
    \item Maximum mass constraint from PSR J0348+0432 reported by ~\cite{Antoniadis_2013}.
    \item Tidal deformability constraint by GW170817 taken from \cite{TheLIGOScientific_2017,Abbott_2017,Abbott_2019}\footnote{ see App. B of~\cite{Montefusco_2025} for a detailed description of how this constraint is implemented.}.
    \item Radius constraints by NICER observations of PSR J0030+0451\cite{Riley_2019,Miller_2019}  from \cite{Riley_2019}, PSR J0740+6620 from \cite{Riley_2021,Miller_2021} from \cite{Riley_2021}, and PSR J0437+4715 from ~\cite{Choudhury_2024,Choudhury_zenodo}. We underline that, while the last constraint is still somewhat controversial, its effect on the posterior distributions in our study has turned out to be negligible when all the other constraints (astrophysical and nuclear) are taken into account.
\end{itemize}

\subsubsection{Nuclear Experimental Constraints}

The nuclear experimental constraint follows \cite{Scurto_2024}, where a Gaussian likelihood model $ P(\{ \mu_i,\sigma_i \}|{\mathbf X})$ is applied to the four NMPs ($\rho_{sat}$, $E_{sat}$, $K_{sat}$, $J_{sym}$), and to the combination 
\begin{equation}
    K_{\tau}=K_{sym}-6L_{sym}-Q_{sat}L_{sym}/K_{sat}\, .
\end{equation}
The Gaussian parameters $\{ \mu_i,\sigma_i \}$, listed in Tab.~\ref{Tab_NMP_constraint}, reflect the expectation and uncertainly of the five parameters taken from \cite{Margueron_2017}, where an indirect constraint on these quantities is obtained by comparing the experimental results with the predictions of different nuclear physics models. 
To implement the constraints on the NMPs, we assume uncorrelated distributions:

\begin{eqnarray}
    P(\{ \mu_i,\sigma_i \}|{\mathbf X})
      =  \prod_{i=1}^{5}  \frac{1}{\sigma_i\sqrt{2\pi}}e^{-\frac{1}{2}(\frac{x_i-\mu_i}{\sigma_i})^2} \, ,
\end{eqnarray}
where $i$ runs over the set~($n_{sat}$, $E_{sat}$, $K_{sat}$, $J_{sym},K_{\tau}$)\footnote{A complete Bayesian inference of the NMPs starting directly from nuclear observables was recently performed  in~\cite{Klausner_2024}.}.

\begin{table}[]
    \centering
    \begin{tabular}{c|cc}
        \hline \hline
        & $\mu$ & $\sigma$\\
        \hline
        $n_{sat}\,$(fm$^{-3})$ & 0.153 & 0.005 \\
        $E_{sat}\,$(MeV)  & -15.8 & 0.3\\
        $K_{sat}\,$(MeV) & 230 & 20\\
        $J_{sym}\,$(MeV) & 32.0 & 2.0\\
        $K_{\tau}\,$(MeV) & -400 & 100\\
        \hline \hline
    \end{tabular}
    \caption{Values of the mean and standard deviation for the NMPs used in the evaluation of $P(\{ \mu_i,\sigma_i \}|{\mathbf X})$, taken from \cite{Margueron_2017}.}
    \label{Tab_NMP_constraint}
\end{table}

\subsubsection{Chiral EFT Constraints}

\begin{figure}
    \centering
    \includegraphics[width=0.9\linewidth]{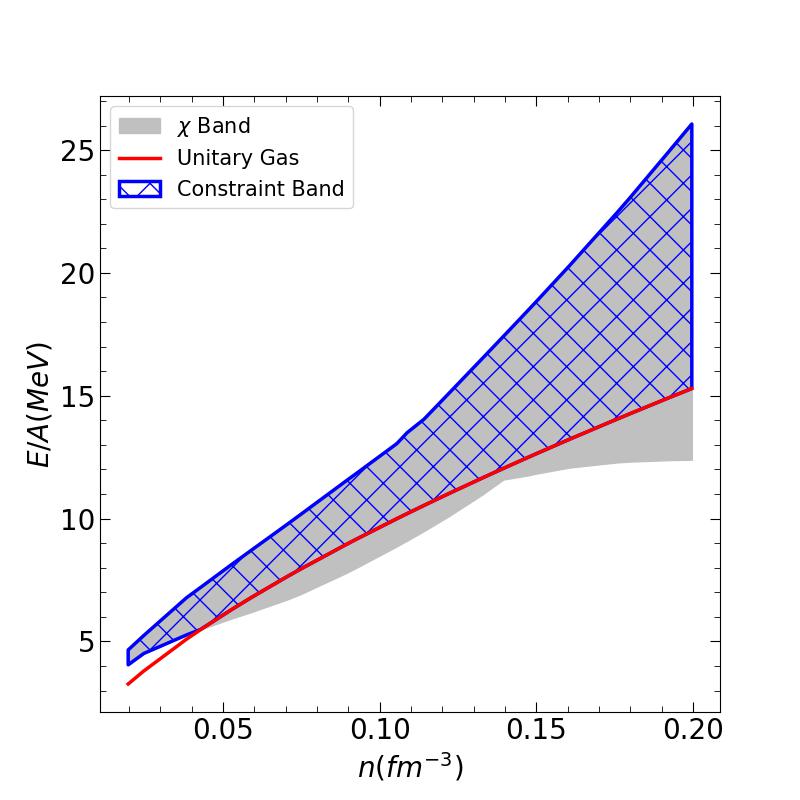}   
    \caption{Energy per particle of pure neutron matter as a function of the density for the $\chi$-EFT constraint used in this work (blue band), of the calculations shown in Fig.~1 of \cite{Huth_2021} (grey band), and of the unitary gas limit~(red solid line). 
    }
    \label{Fig_chiral_band}
\end{figure}

The ab initio constraints on the energy per baryon of neutron matter are taken from \cite{Huth_2021}, which synthesizes results from several previous calculations \citep{Hebeler_2013,Tews_2013,Drischler_2021,Drischler_2019,Lynn_2016}. 
A likelihood model is constructed assuming the energy band defined in \cite{Huth_2021} as a flat 68\% probability distribution, extended according to equation (36) of~\cite{Scurto_2024}. 
For the lower bound of the band, we take the maximum between the lower bound of the band shown in Figure~1 of \cite{Huth_2021} and the energy per baryon of the unitary gas from \cite{Tews:2016jhi} as previously done in~\citet{Montefusco_2025}, see \citet{Sammarruca_2023} for a discussion.
The corresponding energy band used in the constraint is shown in Fig.~\ref{Fig_chiral_band}.

\section{Comparison between Set~A and~Set~B}\label{Sec_4}

\begin{figure*}
       \includegraphics[width=0.9\linewidth,angle=0]{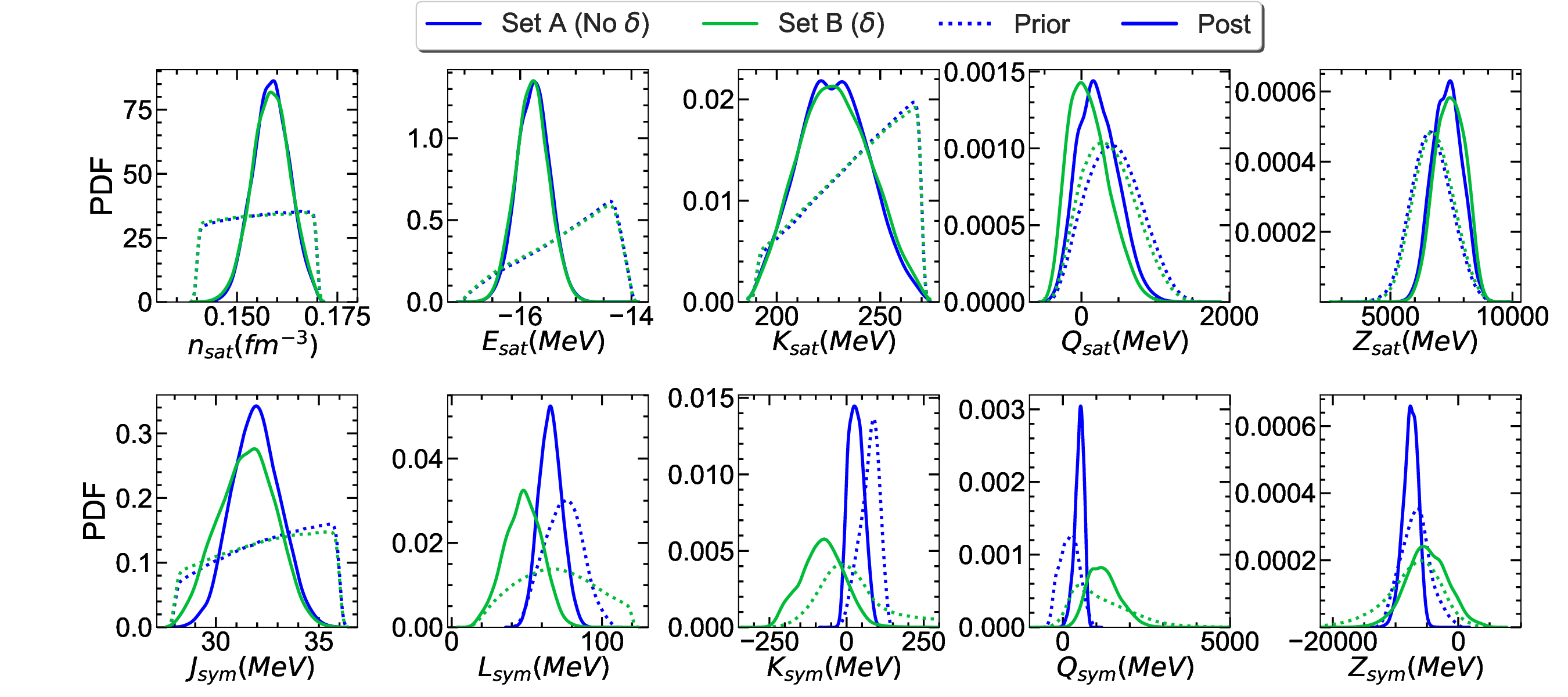}   
    \caption{Marginalized posterior of the NMPs for the prior (dotted) and posterior (solid) distributions for~Set~A~(blue) (without $\delta$ meson) and Set~B~(green) (with $\delta$ meson).}
    \label{Fig_NMP}
\end{figure*}

\begin{figure*}
    \centering
       \includegraphics[width=0.8\linewidth,angle=0]{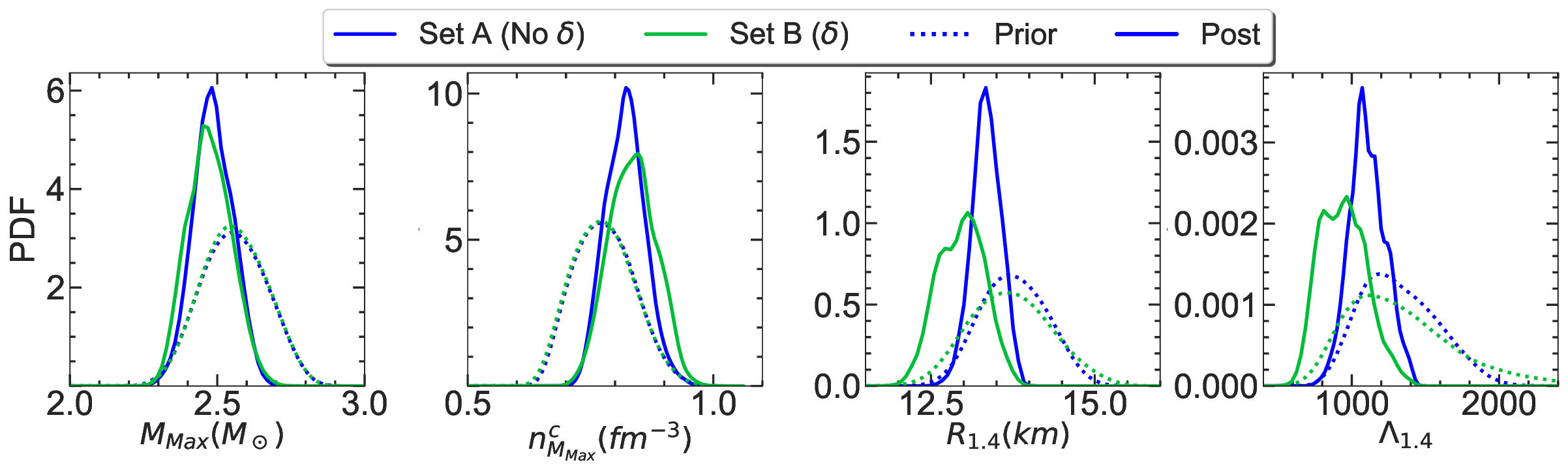}   
    \caption{Marginalized posterior of the maximum mass (first panel), the central density of the most massive star (second panel) and the radius (third panel) and tidal deformability (fourth panel) of a $1.4M_\odot$ NS for the prior (dotted) and posterior (solid) distributions for Set A (blue) and~Set~B~(green).}
    \label{Fig_Astro}
\end{figure*}

\begin{figure}
       \includegraphics[width=0.95\linewidth,angle=0]{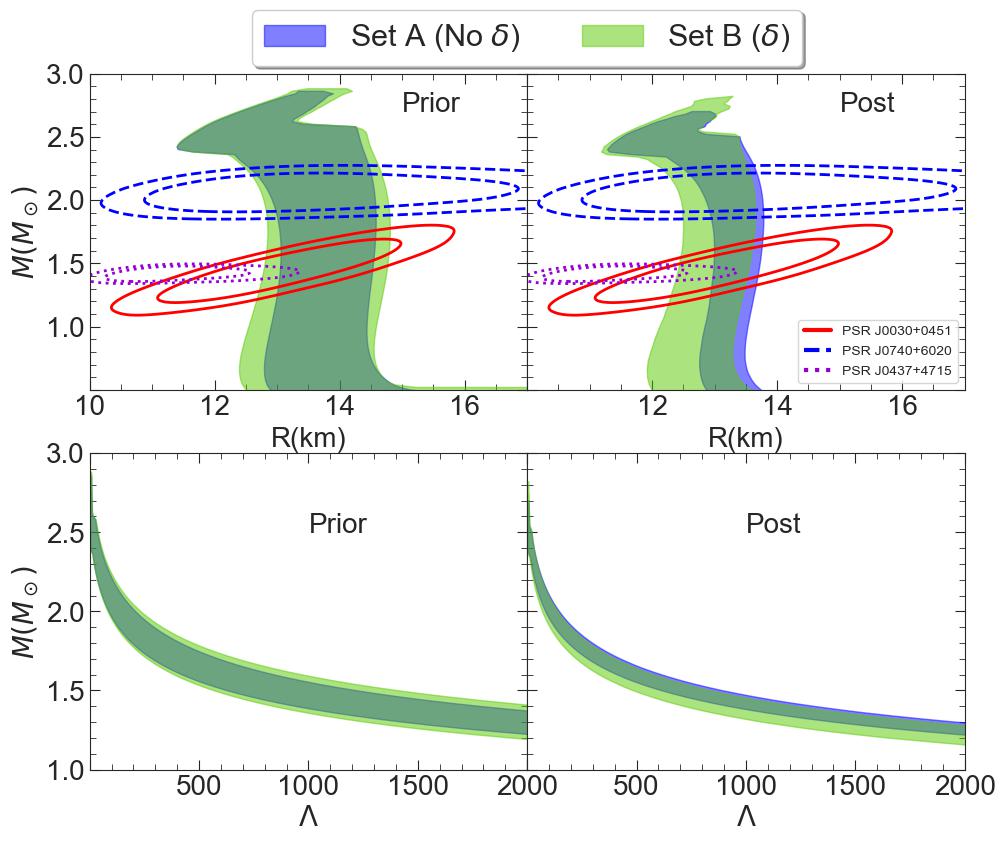}   
    \caption{90\% quantiles for the distribution of the mass as a function of the radius (top panels) and as a function of the tidal deformability (bottom panels), for both the prior (left) and posterior (right) distributions, for Set A (blue) and Set B (green). In the $M$-$R$ plots we also show the 68\% and 90\% quantiles for the three NICER constraints. }
    \label{Fig_MRL}
\end{figure}

\begin{figure}
       \includegraphics[width=0.95\linewidth,angle=0]{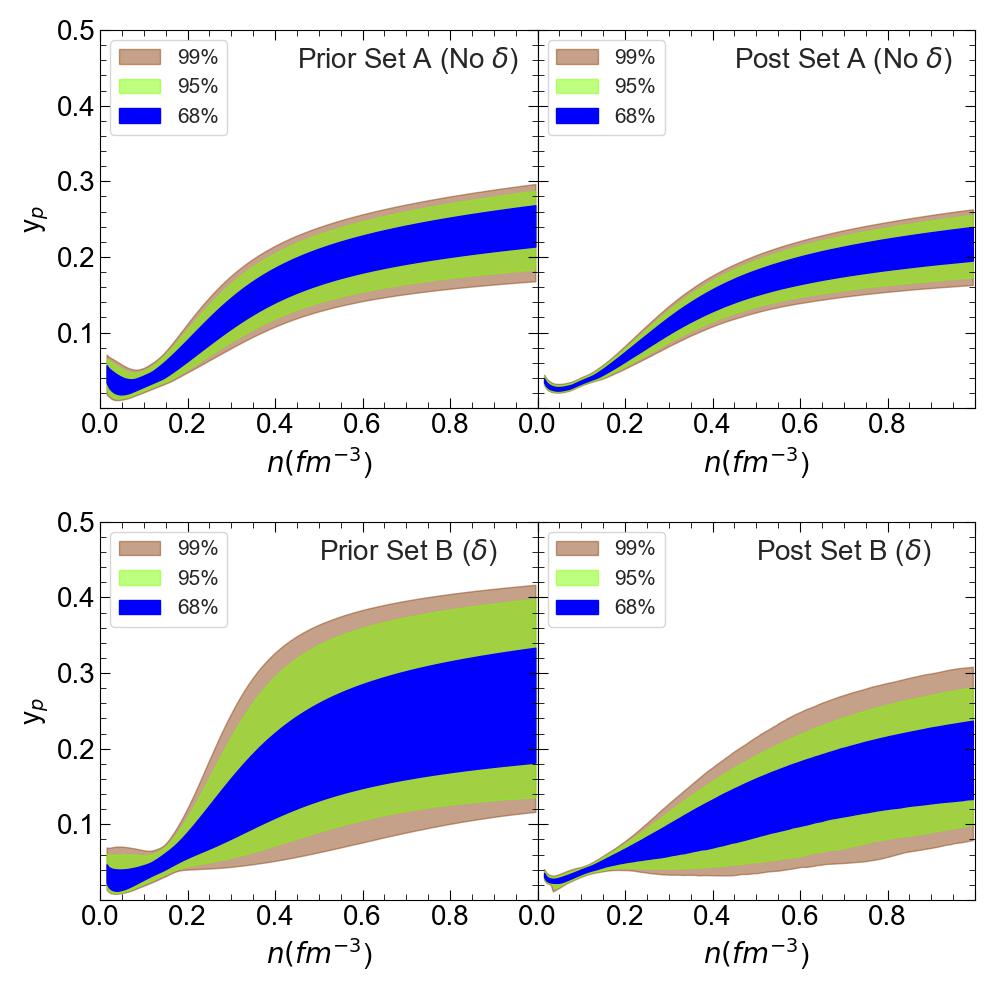}   
    \caption{68\%, 95\% and 99\% quantiles for the distribution of the proton fraction as a function of the baryonic density for the prior (left) and posterior (right) distributions for Set A (top) and Set B (bottom).}.
    \label{Fig_Yp}
\end{figure}

\begin{figure}
    \centering
       \includegraphics[width=0.7\linewidth,angle=0]{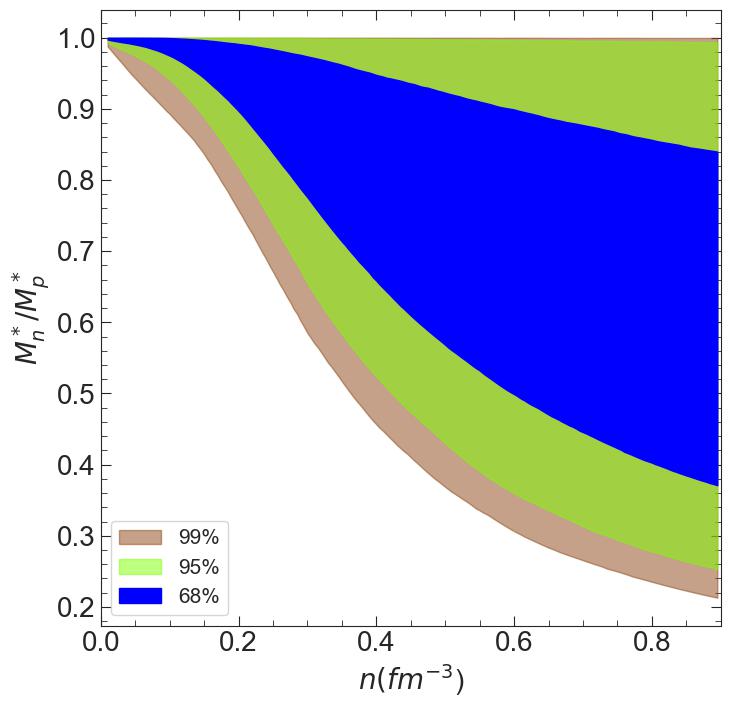}   
    \caption{68\%, 95\% and 99\% quantiles for the posterior distribution of the ratio of the neutron and proton Dirac effective masses as a function of the baryonic density of purely neutron matter for Set B. }
    \label{Fig_Mratio}
\end{figure}

\begin{figure*}
    \centering
       \includegraphics[width=0.7\linewidth,angle=0]{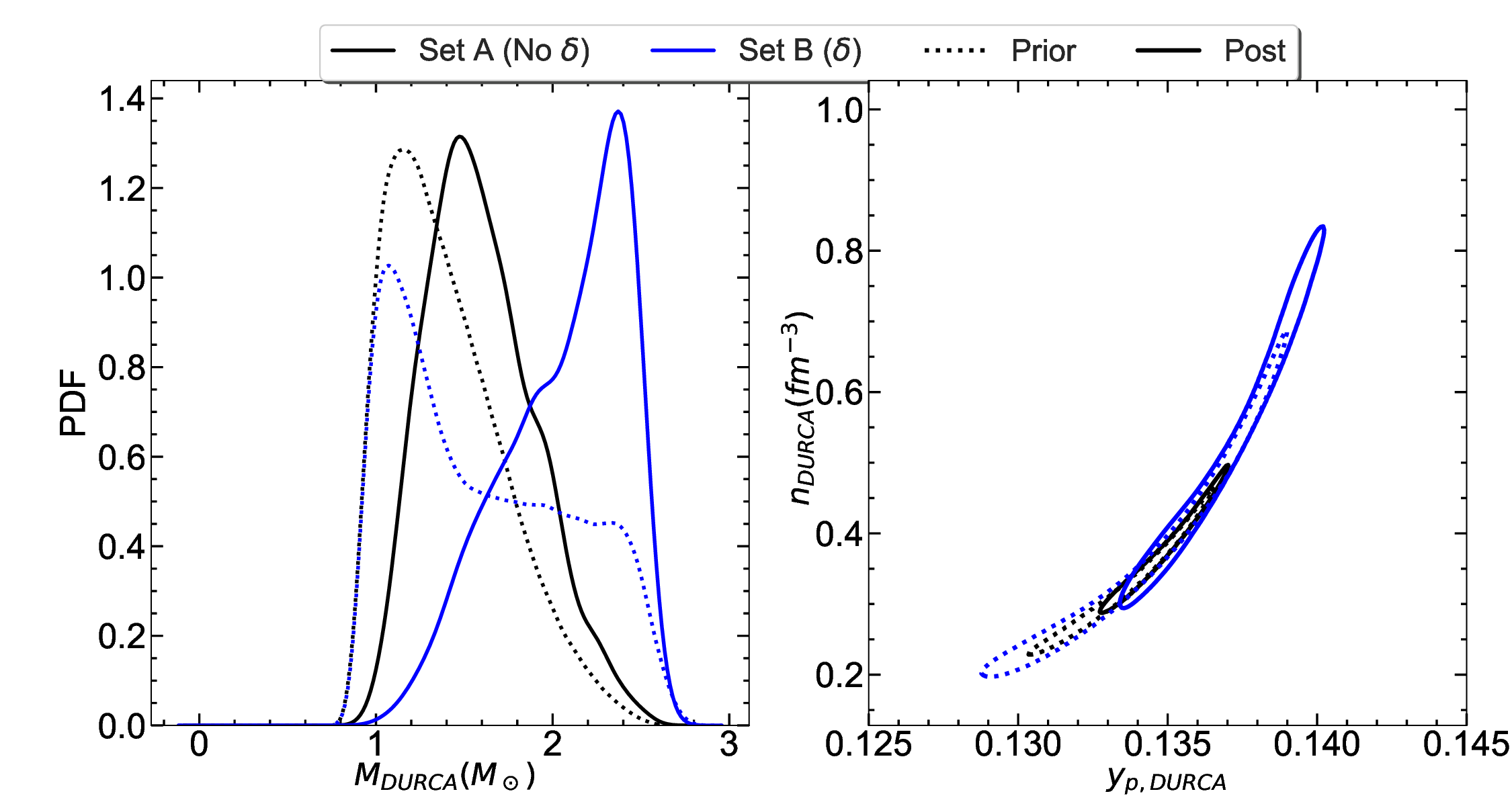}   
    \caption{Marginalized posterior of the mass at which the dUrca threshold is met (left) and the 2-D marginalized posterior for the baryonic density and proton fraction for the same threshold, for the prior (dotted) and posterior (solid) distributions for Set A (black) and Set B (blue)}.
    \label{Fig_DURCA}
\end{figure*}

In this section, we compare the results obtained for Set A and Set B. Figure~\ref{Fig_NMP} presents the prior and posterior distributions of the NMPs for both sets. The distributions of the isoscalar sector parameters are nearly identical, as expected due to the isovectorial nature of the $\delta$ meson. However, a notable difference emerges in the symmetry NMPs, particularly in the distributions of $L_{sym}$ and~$K_{sym}$. 

Specifically, we observe that the presence of the $\delta$ meson in Set B leads to broader distributions. Additionally, for both $L_{sym}$ and $K_{sym}$, the peak of the distribution shifts to lower values compared to Set A. Notably, the results for Set B align well with those reported in \cite{Scurto_2024}, demonstrating that the inclusion of the additional meson provides the model with a degree of flexibility previously achieved through a more complex density functional for the couplings. 
Furthermore, in agreement with \cite{Scurto_2024}, we find that the constraints imposed in the posterior calculation tend to prefer lower values for both $L_{sym}$ and $K_{sym}$.

In Fig.~\ref{Fig_Astro}, we present the prior and posterior distributions for key astrophysical observables: the mass and central density of the most massive star, as well as the radius and tidal deformability of a $1.4M_\odot$ star. The figure illustrates that the distributions of the parameters related to the most massive star remain largely unaffected by the presence of the isospin mass splitting.
This outcome also aligns with the findings of \cite{Scurto_2024}, which indicate that these quantities are more strongly correlated with the isoscalar sector, which, in turn, is not significantly influenced by the $\delta$ meson. Conversely, the properties of the $1.4M_\odot$ star exhibit noticeable modifications due to the additional meson, with broader distributions for both radius and deformability, allowing lower values in the case of Set B. The significance of this result is further highlighted in Fig.~\ref{Fig_MRL}, where we depict the 90\% quantiles of the prior and posterior distributions for the $M$-$R$ and $M$-$\Lambda$ relations for both sets. In the case of the $M$-$R$ relation, we also overlay the three NICER constraints. This demonstrates that the models featuring a lower radius, present in Set B, play a crucial role in ensuring compatibility with the constraint from PSR~J0437+4715, which favours more compact objects (difficult to generate within our RMF formalism). Furthermore, the figure illustrates that the priors for Set A are entirely enclosed within those of Set B, reinforcing the notion that the $\delta$ meson enhances the model's flexibility.

Fig.~\ref{Fig_Yp} displays the distribution of the proton fraction $y_p=n_p/(n_n+n_p)$ as a function of density for both sets. A direct comparison with the results of \cite{Scurto_2024} immediately reveals the impact of changes in the EoS model. Specifically, for Set A, the absence of the cubic term in the density-dependent couplings - see equation~\eqref{gabc} - results in a narrower distribution of $y_p$. This occurs due to the lack of models exhibiting a sharp increase in proton fraction around $0.4$ fm$^{-3}$, a feature observed in our previous study. In contrast, when considering Set B, we find that the introduction of the $\delta$ meson not only restores the broadness of the distribution seen previously but also smooths the proton fraction behaviour. Additionally, the distribution extends symmetrically toward both higher and lower values of $y_p$ at higher densities, yielding models with $y_p$ below $0.1$ at high densities, that were absent in the previous study.

The existence of models with such low $y_p$ can be explained by examining Fig.~\ref{Fig_Mratio}, which presents the ratio of neutron and proton Dirac masses as a function of density for neutron matter in Set B. The figure reveals that, due to the presence of the $\delta$ meson, the effective Dirac mass of the proton can become nearly two and a half times larger for the 68\% and up to five times larger than that of the neutron for the 99\% quantile.
This substantial mass difference favours neutron-rich matter, thereby explaining the very low proton fractions observed in the previous plot.

Finally, in Fig.~\ref{Fig_DURCA}, we present the distributions for the mass, density, and proton fraction at which the threshold for the dUrca process is reached. The dUrca threshold for the proton fraction is given by~\cite{Klahn2006}

as
\begin{equation}
    y_{p,dUrca}=\frac{1}{ 1+\left(1+x_e^{1/3} \right)^3 }
    \label{Eq_DURCA}
\end{equation}
where $x_e=n_e/(n_e+n_\mu)$ represents the leptonic electron fraction. The threshold density $n_{dUrca}$ is defined via $y_p(n_{dUrca})=y_{p,dUrca}$, while the threshold mass $M_{dUrca}$ corresponds to the mass of a star with central density~$n_{dUrca}$.

We observe that the distributions for Set A are significantly narrower compared to those of Set B. Additionally, for both sets, the posterior constraint shifts the peak of the mass distribution toward higher values. This effect is more pronounced in Set B due to the narrower distribution of Set A, which results in a lack of models with high $M_{dUrca}$. The shift in threshold mass is directly linked to the corresponding shift in the joint distribution of $n_{dUrca}$ and $y_{p,dUrca}$. Similarly, we find that the posterior distribution shifts toward higher density values, a trend that can be traced back to Fig.~\ref{Fig_Yp}: in agreement with \cite{Scurto_2024}, the constraints applied to obtain the posterior (particularly the $\chi$-EFT constraint) tend to disfavour models exhibiting a steep increase in proton fraction.

\begin{figure*}
    \centering
       \includegraphics[width=0.8\linewidth,angle=0]{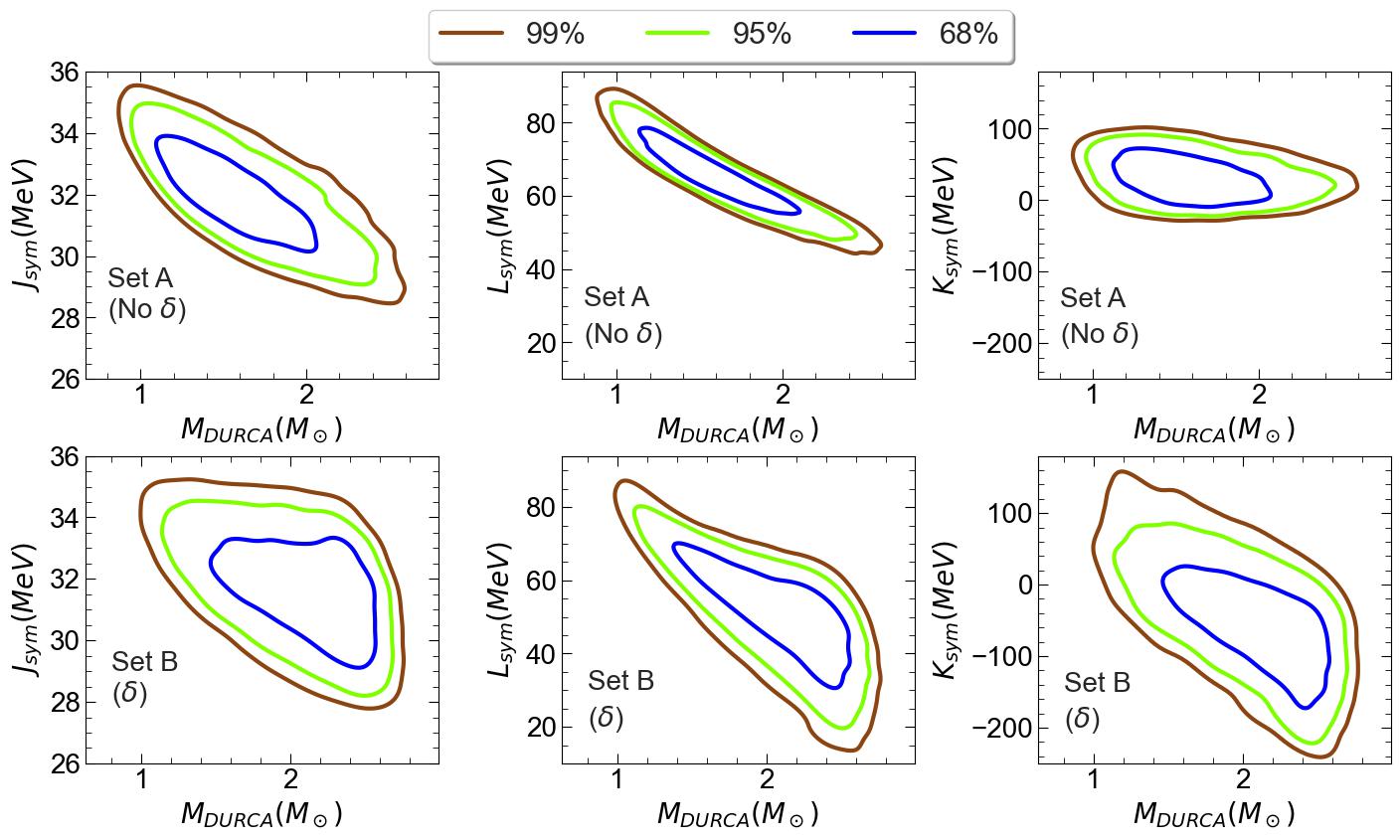}   
    \caption{68\% (blue) 95\% (green) and 99\% (brown) quantiles for the 2-D posterior distributions of $J_{sym}$ (left), $L_{sym}$ (center) and $K_{sym}$ (right) with $M_{dUrca}$ for Set A (top row) and Set B (bottom row).}
    \label{Fig_JLK}
\end{figure*}

\begin{figure*}
    \centering       \includegraphics[width=0.8\linewidth,angle=0]{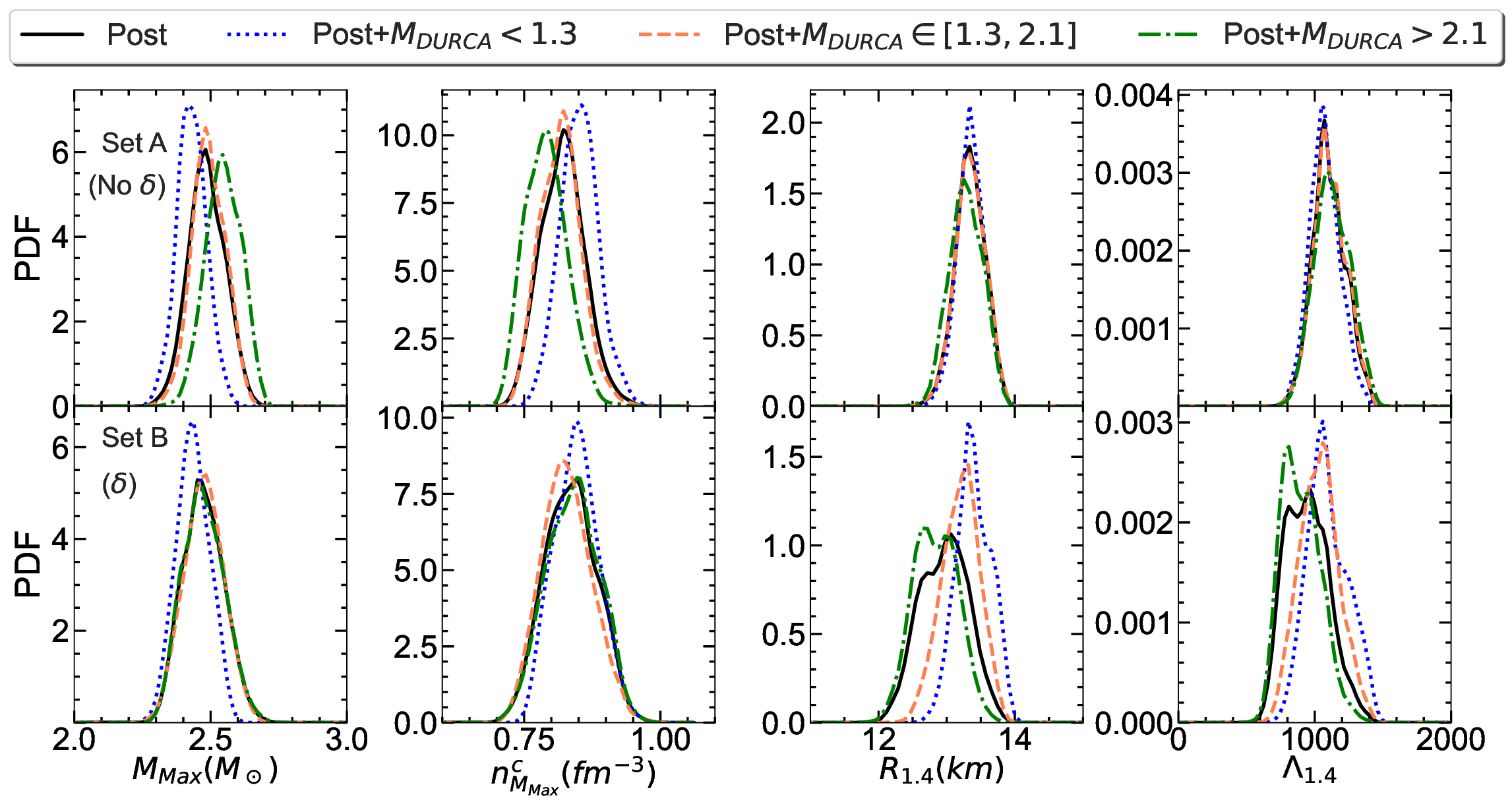}   
    \caption{Marginalized posterior of the astrophysical observables shown in Fig.\ref{Fig_Astro} for Set A (top row) and Set B (bottom row). We show the results for the posterior without any assumption on the dUrca threshold (black solid) as well as the ones including the three assumptions on the threshold "Always" ($M_{dUrca}<1.3M_\odot$), "Sometimes" ($1.3M_\odot\leq M_{dUrca}\leq 2.1M_\odot$) and "Never" ($M_{dUrca}>2.1M_\odot$) (blue dotted, orange dashed and green dash-dotted respectively).}.
    \label{Fig_Astro_D}
\end{figure*}

\section{Direct Urca Constraint}\label{Sec_5}

Recent studies on NS cooling by \cite{Marino_2024} reinforce the findings of previous works \citep{Potekhin_2018,Potekhin_2023}, highlighting the necessity of fast cooling processes to account for the observed young but cold neutron stars. 
Based on this premise, we assume that dUrca processes are essential and treat this as a hypothetical constraint. Following \cite{Margueron_2018}, we compare the predictions obtained under different assumptions regarding the onset of the dUrca threshold.

We consider three possible scenarios. In the first, labelled "Always," we assume that the dUrca threshold is already met at low masses ($M_{dUrca}<1.3M_\odot$). In the second, termed "Sometimes," the threshold falls within the observed mass range of NSs ($1.3M_\odot\leq M_{dUrca}\leq 2.1M_\odot$), as suggested by \cite{Marino_2024}. Finally, in the "Never" scenario, the threshold is either reached at masses exceeding the observed range ($M_{dUrca}>2.1M_\odot$) or is not met at all.

While, as previously shown in \cite{Margueron_2018}, the hypothesis on the dUrca does not significantly affect the distribution of the isoscalar sector of the NMPs, in the isovector sector, the dUrca constraint has a significantly stronger effect on the NMPs.  
For both sets, a lower dUrca threshold leads to higher values of $J_{sym}$, $L_{sym}$, and $K_{sym}$, as it can be seen in Fig.~\ref{Fig_JLK}. This behaviour aligns with the trends seen in Fig.~\ref{Fig_DURCA}, where the posterior distribution in Set A (black line) is closer to the "Always" and "Sometimes" scenarios, whereas in Set B, it aligns more closely with the "Sometimes" and "Never" cases. This difference arises because, in Set A, the posterior distribution of $M_{dUrca}$ is shifted toward lower values, with a limited number of models exhibiting a very high threshold mass. Conversely, in Set B, the mass distribution is shifted toward higher values, with fewer models having a low threshold mass.

Finally, in Fig.~\ref{Fig_Astro_D}, we present the distributions of astrophysical observables. We observe that the quantities related to the most massive star are not strongly affected by the dUrca constraint.
This can be attributed to the fact that, as previously discussed, these quantities are more strongly correlated with the NMPs in the isoscalar sector, which, is not significantly influenced by the dUrca constraint. 
Conversely, the observables associated with the $1.4M_\odot$ star are more constrained in Set B. Notably, we find that the secondary peaks in the posterior distribution, appearing at lower radii and tidal deformabilities, are almost entirely associated with models from the "Never" scenario and vanish in the "Always" and "Sometimes" scenarios. This is consistent with the fact that Set A lacks both models with very high dUrca threshold masses and models with low radii and tidal deformabilities for the $1.4M_\odot$ star.

\begin{figure*}
    \centering       \includegraphics[width=0.7\linewidth,angle=0]{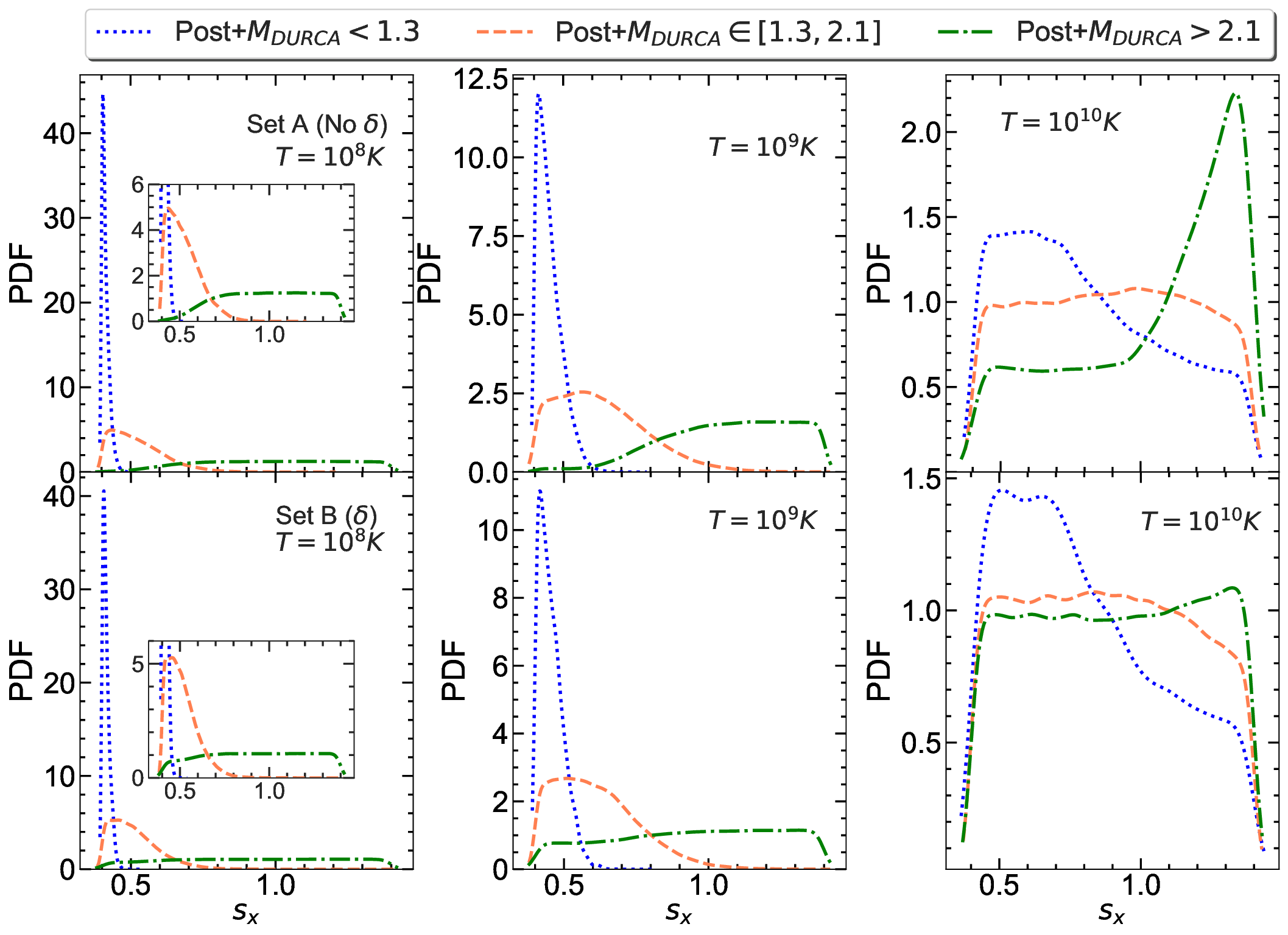}   
    \caption{Marginalized posterior of the $s_x$ parameter for Set A (top row) and Set B (bottom row) and for the three considered star temperatures, namely $T=10^8K$ (left panels),$T=10^9K$ (central panels) and $T=10^{10}K$ (right panels). We show the results for the three assumptions on the threshold "Always" ($M_{dUrca}<1.3M_\odot$), "Sometimes" ($1.3M_\odot\leq M_{dUrca}\leq 2.1M_\odot$) and "Never" ($M_{dUrca}>2.1M_\odot$) (blue dotted, orange dashed and green dash-dotted respectively).}.
    \label{Fig_Sx_D}
\end{figure*}

\begin{figure*}
    \centering       \includegraphics[width=0.7\linewidth,angle=0]{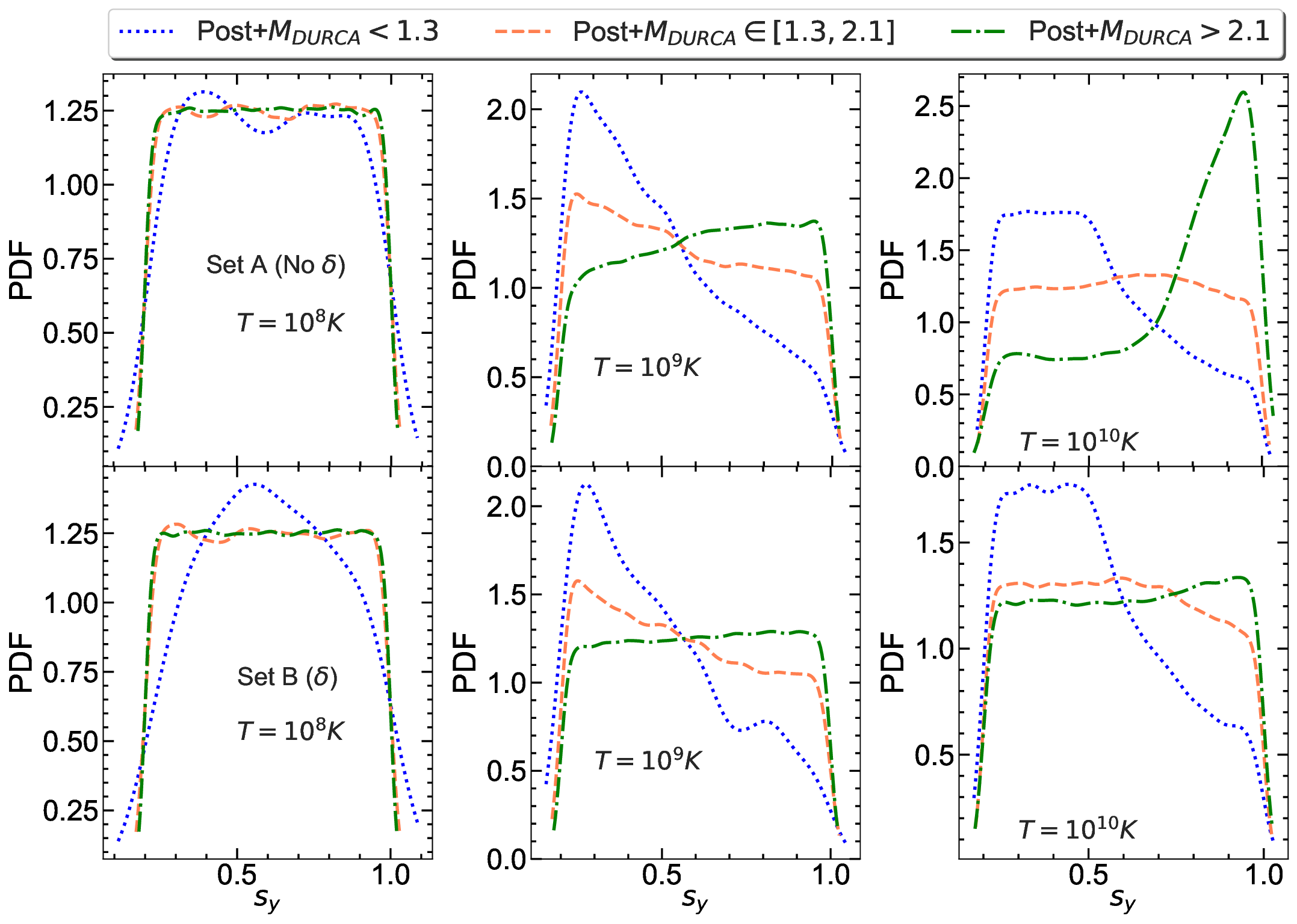}   
    \caption{Marginalized posterior of the $s_y$ parameter for Set A (top row) and Set B (bottom row) and for the three considered star temperatures, namely $T=10^8K$ (left panels),$T=10^9K$ (central panels) and $T=10^{10}K$ (right panels). We show the results for the three assumptions on the threshold "Always" ($M_{dUrca}<1.3M_\odot$), "Sometimes" ($1.3M_\odot\leq M_{dUrca}\leq 2.1M_\odot$) and "Never" ($M_{dUrca}>2.1M_\odot$) (blue dotted, orange dashed and green dash-dotted respectively).}.
    \label{Fig_Sy_D}
\end{figure*}

\begin{figure}
    \centering       \includegraphics[width=0.98\linewidth,angle=0]{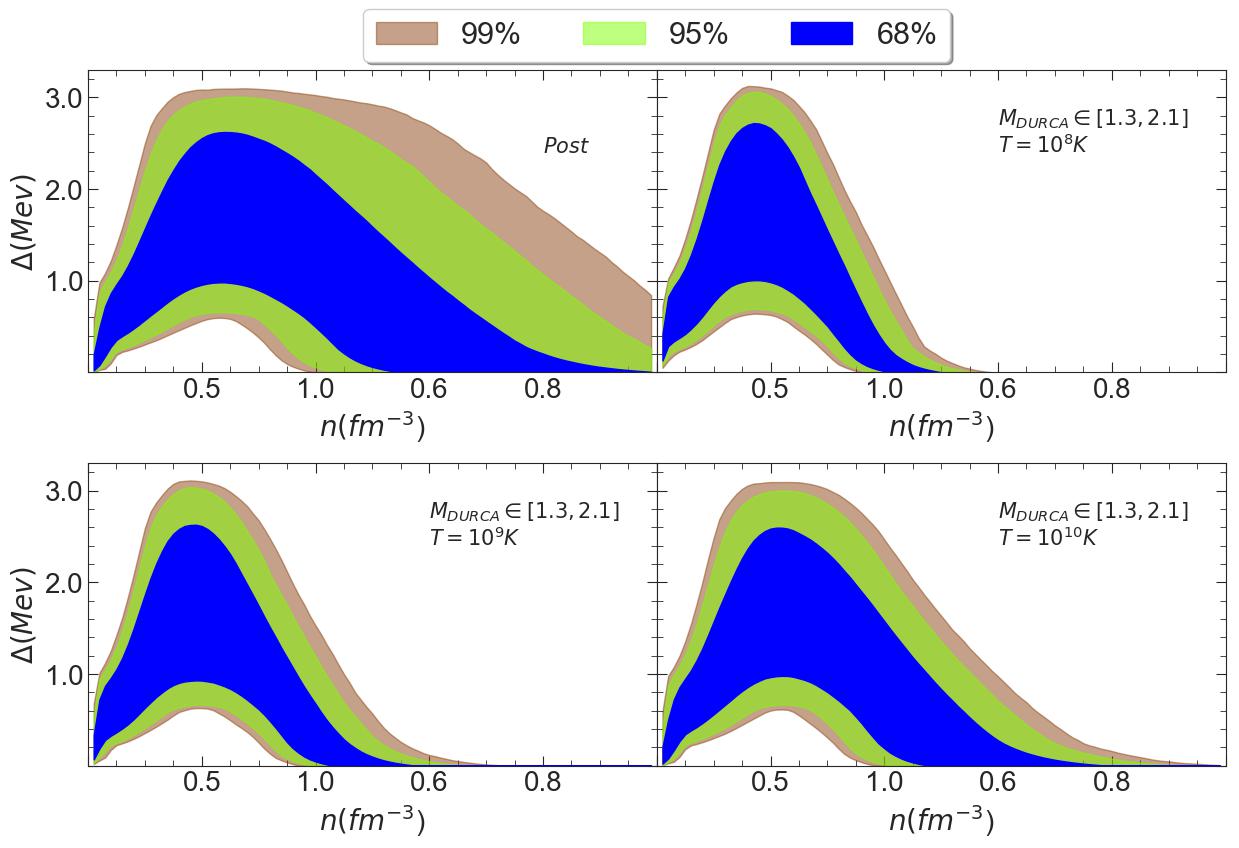}  
    \caption{68\% (blue) 95\% (green) and 99\% (brown) quantiles for the posterior distributions of the pairing gap as a function of the baryonic density of beta-equilibrated matter for Set B. We show the results for the posterior without assumptions on the dUrca threshold (top left), as well as when the "Sometimes" ($1.3M_\odot\leq M_{dUrca}\leq 2.1M_\odot$) assumption for the dUrca threshold is made for a star at $T=10^8K$ (top right), $T=10^9K$ (bottom left) and $T=10^{10}K$ (bottom right).}.
    \label{Fig_Gap_D2}
\end{figure}

\subsection{Introduction of the 1S0 Proton Pairing Gap}

Superfluidity plays a crucial role in the cooling of neutron stars, as discussed for instance in \cite{Yakovlev1999}, see \citet{Potekhin2015review} for a review. In particular, proton superfluidity in the core suppresses the dUrca process, as shown by \cite{Wei_2019}. Given the strong theoretical uncertainty on the effect of the medium polarization on the  gap discussed by \cite{Lombardo_2000,Drissi_2022,Pavlou_2017,Benhar_2017,Gandolfi_2022}, we introduce a simplified model for the 1S0 proton pairing gap $\Delta$ following \cite{Wei_2020,Das_2024}, which is a rescaled version of the BCS pairing gap as calculated in ~\cite{Lombardo_2000}:
\begin{equation} 
\Delta(n_p,s_y,s_x)=s_y\Delta_{BCS}(n_p/s_x) \, , 
\label{Eq_gap} 
\end{equation}
where $\Delta_{BCS}$ represents the BCS pairing gap, $n_p$ is the proton density and $s_x$ and $s_y$ are free scaling parameters that encode the uncertainty on the polarization corrections to the BCS result. 
The gap is implemented by generating 100 pairs of $s_x$ and $s_y$ values for each model in both sets, drawing them from uniform distributions with limits $0.4\leq s_x\leq 1.4$ and $0.2\leq s_y\leq 1.0$, taken from~\cite{Das_2024}.

In principle, enhanced cooling via dUrca should be accounted for throughout the entire cooling history of a neutron star. However, properly assessing the relevance of different cooling processes would require coupling our approach with a full cooling code, see \cite{Potekhin2015review}, which is beyond the scope of this study.
Studies on the thermal evolution of NSs by \cite{Yakovlev_2004} and \cite{Page_2004} indicate that they rapidly cool to surface temperatures of approximately $T=10^{6.5}\,$K, which, according to thermal profile studies of \cite{Gudmundsson_1983}, corresponds to core temperatures around $T=10^{9}\,$K. 

Building on these findings, to gain at least a qualitative understanding of the constraints that potential evidence of enhanced cooling could impose on the underlying microphysics, we consider three schematic stellar temperature scenarios: a hot star ($T=10^{10}\,$K), a warm star ($T=10^{9}\,$K), and a cold star~($T=10^{8}\,$K).
For each case, we assume that the dUrca process is suppressed as long as the gap remains above a threshold that was estimated by \cite{BCS_1957}, see also \cite{Chamel_2008}: 
\begin{equation} 
\Delta_T=1.56k_BT \, , 
\label{Eq_gap_T} 
\end{equation}
where $k_B$ is the Boltzmann constant. At each temperature, the density for the dUrca threshold is determined as the maximum between $n_{dUrca}$ (as defined in the previous section) and $n_\Delta$, where $n_\Delta$ is  the density at which the proton density in the gap equation~(\ref{Eq_gap})  satisfies the above condition, Eq.~(\ref{Eq_gap_T}). 

By applying the same three assumptions for the dUrca threshold as in the previous section, we obtain more realistic estimates for the posterior EoS distribution while simultaneously imposing constraints on the pairing gap.
Due to the large uncertainty on the gap, the distributions for the NMPs turn out to not be significantly modified with respect to what is shown in Fig. \ref{Fig_NMP}. On the other hand, the gap itself is significantly constrained by the three scenarios considered for the dUrca process. In Fig.~\ref{Fig_Sx_D} and \ref{Fig_Sy_D}, we present the resulting distributions for $s_x$ and $s_y$, respectively.
For $s_x$, we observe that in the $T=10^{8}\,$K and $T=10^9\,$K cases, both the "Always" and "Sometimes" dUrca scenarios impose strong constraints, favouring lower values of $s_x$ as expected. The results for Set A and Set B are largely similar, with the primary difference being the shape of the "Never" distribution at the highest temperature. This difference arises because Set A contains fewer models with high $n_{dUrca}$, meaning that in this set, the "Never" distribution is dominated by models with higher $n_\Delta$, which corresponds to higher values of~$s_x$.

For $s_y$, the $T=10^{8}~$K case does not provide any significant constraints, as the density at which the gap vanishes is independent of $s_y$. At the intermediate temperature, the constraint on $s_y$ is noticeably weaker compared to that on $s_x$, with the entire parameter range being included in the distributions for all three dUrca scenarios. The peak observed in the "Never" distribution at the highest temperature follows the same pattern as in the $s_x$ distribution.

Finally, in Fig.~\ref{Fig_Gap_D2}, we show the posterior distribution of the pairing gap as a function of baryonic density, comparing results without any dUrca constraint to those under the "Sometimes" dUrca threshold scenario. The three stellar temperatures discussed earlier are considered.
We show only the results for Set B since, while the distribution in the case without dUrca constraint appears to be narrower in the case of Set A, as soon as the dUrca constraint is considered, the distributions for Set A and Set B appear to be almost undistinguishable. In the figure we can notice that, when the dUrca constraint is imposed, we observe a considerable narrowing of the distribution along the x-axis compared to the unconstrained case. As expected, this effect is more pronounced at lower stellar temperatures, consistent with Fig.~\ref{Fig_Sx_D}. Meanwhile, the weak constraint on $s_y$, as seen in Fig.~\ref{Fig_Sy_D}, explains why the distribution remains largely unchanged along the y-axis under the dUrca hypothesis.

\section{Conclusions}
\label{Sec_6}

We analyzed the impact of the considerable uncertainty on  the neutron-proton effective mass splitting,  by the introduction of an additional scalar-isovector meson coupling in the RMF formalism, the $\delta$ meson. Following \cite{Scurto_2024}, we conducted a Bayesian analysis comparing results obtained with and without this additional meson. Our findings indicate that the $\delta$ meson significantly influences both nuclear matter properties and NS observables. Specifically, we demonstrated that the inclusion of the $\delta$ meson substantially broadens the distribution of NMPs in the isovector sector, shifting the mean values of both $L_{sym}$ and $K_{sym}$ to lower values.

Regarding NS observables, we found that while the properties of the most massive star remain largely unchanged, the presence of the additional meson results in broader distributions for both the radius and tidal deformability of stars with masses around~$1.4 M_\odot$.

Moreover, we showed that when the $\delta$ meson is included, the proton fraction distribution in beta-equilibrated matter becomes significantly broader at higher densities, allowing for models with $y_p<0.1$ at 1 fm$^{-3}$. This effect extends to the distribution of the dUrca process threshold, which increases notably in the set including $\delta$. Additionally, the peak of the mass threshold distribution shifts to higher values.

The substantial impact of the $\delta$ meson on the dUrca threshold motivated us to introduce a hypothetical constraint on the mass threshold for this process. Comparing three different scenarios against this constraint revealed that different assumptions about the threshold's onset yield varying predictions for the mean values of NMPs in the isovector sector. 

Finally, we introduced the simplified but flexible model for the $^1S_0$ proton pairing gap, as assumed in~\cite{Das_2024}. We demonstrated that the dUrca threshold constraint can be used to make predictions about about the in-medium effects on the pairing gap, which are known
to modify the simple prediction of the BCS theory, but are not fully assessed. In particular, we showed that assuming the dUrca threshold to lie within the range $1.3M_\odot \leq M_{dUrca} \leq 2.1M_\odot$ strongly constrains the $s_x$ parameter, thereby limiting the possible density interval associated to the superconducting phase.
This has potentially interesting consequences for the cooling history of neutron stars, that we plan to investigate with more realistic cooling simulations in a future work.

\section{Acknowledgements}

The authors acknowledge the Laboratory for Advanced Computing at the University of Coimbra for providing computing resources that have contributed to the research results reported within this paper. URL: https://www.uc.pt/lca. L.S. acknowledges the PhD grant 2021.08779.BD (FCT, Portugal)  with DOI identifier 10.54499/2021.08779.BD. Partial support from the IN2P3 Master Project NewMAC and the ANR project GW-HNS (ANR-22-CE31-0001-01).
This work was also partially supported by national funds from FCT (Fundação para a Ciência e a Tecnologia, I.P, Portugal) under projects UIDB/04564/2020 and UIDP/04564/2020, with DOI identifiers 10.54499/UIDB/04564/2020 and 10.54499/UIDP/04564/2020, respectively, and the project 2022.06460.PTDC with the associated DOI identifier 10.54499/2022.06460.PTDC. H.P. acknowledges the grant 2022.03966.CEECIND (FCT, Portugal) with DOI identifier 10.54499/2022.03966.CEECIND/CP1714/CT0004.

\bibliographystyle{aa}

\begin{appendix}

\section{Relativistic Mean Field Formalism Details} \label{App_RMF}

In this appendix, we provide a detailed exposition of the formalism employed in the construction of the EoS models used in this study.
We start by showing the terms in Eq.~\eqref{Eq_1}. The second term in Eq.~\eqref{Eq_1} represents an ideal Fermi gas of muons and electrons, 
\begin{equation}
    \mathcal{L}_L=\sum_{l=e,\mu}\bar{\psi}_l\left[
    i\gamma_\mu \partial^\mu-m_l \right] \psi_l \, .
\end{equation}
Finally, the mesonic terms in the Lagrangian are given by the standard expressions
\begin{align}
    \mathcal{L}_\sigma & = \frac{1}{2}\bigg(\partial_\mu\phi\partial^\mu\phi-m_\sigma^2\phi^2\bigg) \, ,
    \\
    \mathcal{L}_\delta & = \frac{1}{2}\bigg(\partial_\mu\boldsymbol{\delta}\cdot\partial^\mu\boldsymbol{\delta}-m_\delta^2\boldsymbol{\delta}\cdot\boldsymbol{\delta}\bigg) \, ,
    \\
    \mathcal{L}_\omega & = -\frac{1}{4}\Omega_{\mu\nu}\Omega^{\mu\nu}+\frac{1}{2}m_\omega^2V_\mu V^\mu\, ,
    \\
    \mathcal{L}_\rho & = -\frac{1}{4}\mathbf{B}_{\mu\nu}\cdot\mathbf{B}^{\mu\nu}+\frac{1}{2}m_\rho^2\mathbf{b}_\mu\cdot \mathbf{b}^\mu \, ,
\end{align}
with the tensors written as
\begin{align}
\Omega_{\mu\nu} &= \partial_\mu V_\nu - \partial_\nu V_\mu \, , 
\\
\mathbf{B}_{\mu\nu} &= \partial_\mu \mathbf{b}_\nu - \partial_\nu \mathbf{b}_\mu - g_\rho \left(\mathbf{b}_\mu \times \mathbf{b}_\nu \right)\, .
\end{align}
In the mean-field approximation, the mesonic fields are replaced by their ground state expectation value, so that for each meson only the third vector component and the zeroth isovector component are different from zero. These components must satisfy the classical Euler-Lagrange equations: 
\begin{align}
    m_\sigma^2 \, \phi_0 &= g_\sigma \, n_s \\
    \label{EL_delta}
    m_\sigma^2 \, \delta_3 &= g_\delta \, n_{s3} \\
    m_\omega^2 \, V_0 &= g_\omega \, n \\
    m_\rho^2 \, b_{03} &= \frac{g_\rho}{2} \, n_3 \, .
\end{align}
In the above equations, $n=n_{n}+n_{p}$ is the total baryon number density, $n_s=n_{sn}+n_{sp}$ is the total scalar density, while $n_3=n_p-n_n$ and $n_{s3}=n_{sp}-n_{sn}$. The two scalar densities are obtained as 
\begin{align}
    n_{si}=\frac{1}{\pi^2}\int_0^{k_{Fi}} \!\! dk \,
    \frac{M_i^* k^2}{\sqrt{k^2+M_i^{*2}}} \, ,
\end{align}
$k_{Fi}$ being the Fermi momentum of neutrons ($i=n$) and protons ($i=p$).
The nucleonic chemical potentials are given by 
\begin{align}
\mu_i=\sqrt{k^2_{Fi}+M_i^{*2}}+g_\omega\omega_0
\pm \frac{1}{2}g_\rho b_{03}+\Sigma_R 
\end{align}    
where $i=p,n$, the plus sign is for protons and the minus is for neutrons and where $\Sigma_R$ is the rearrangement term arising from the density dependence of the couplings:
\begin{align}
    \Sigma_R(\rho)=\frac{\partial g_\omega}{\partial\rho}\omega_0n+\frac{1}{2}\frac{\partial g_\rho}{\partial \rho}b_{03} n_3-\frac{\partial g_\sigma}{\partial \rho}\phi_0n_s-\frac{\partial g_\delta}{\partial \rho}\delta_3n_{s3} \, .
\end{align}
Finally, the energy density and pressure are given by 
\begin{equation}
\mathcal{E}(n,n_3)= \sum_{i=n,p,e,\mu}   \mathcal{E}_{kin,i}+\mathcal{E}_{field}   \, , 
\end{equation}
and 
\begin{equation}
P = \sum_{i=n,p,e,\mu} P_{kin,i} + P_{field}
\end{equation}
where 
\begin{align}
   \mathcal{E}_{kin,i}&= \frac{1}{\pi^2}\int_0^{k_{Fi}} dk\,k^2 \sqrt{k^2+M_i^{*2}}  
   \\
  \mathcal{E}_{field} &=\frac 12 m_\omega^2 \omega_0^2 + \frac 12 m_\rho^2 b_{30}^2 + \frac 12 m_\sigma^2 \phi_0^2 + \frac 12 m_\delta^2 \delta_3^2  
\end{align}
and
\begin{align}
  P_{kin,i}&= \frac{1}{\pi^2}\int_0^{k_{Fi}} dk\,\frac{k^4} {\sqrt{k^2+M_i^{*2}}} 
  \\
  P_{field}&=\frac 12 m_\omega^2 \omega_0^2 + \frac 12 m_\rho^2 b_{30}^2  
  \\ \nonumber
  &- \frac 12 m_\sigma^2 \phi_0^2 - \frac 12 m_\delta^2 \delta_3^2 +n\Sigma_R \, .
\end{align}
Charge neutrality $\rho_p=\rho_e+\rho_\mu$ and the weak equilibrium conditions
\begin{equation}
 \mu_n-\mu_p=\mu_e \; \qquad \qquad \; \mu_e=\mu_\mu   
\end{equation}
complete the equations set for homogeneous matter in the core.

For the non-homogeneous crust, we follow the same prescription as in \cite{Scurto_2024}. In the outer crust, the BSK22 equation of state \cite{Pearson_2018} is adopted for all equations of state in our sets, while in the inner crust, a CLD approximation consistent with the RMF formalism used for the core is employed~\cite{Pais_2015,Scurto_2023}.
In this CLD formalism, each Wigner-Seitz cell is composed of a constant electron background of energy density $\mathcal{E}_L$, a high-density ("cluster") part, labelled $I$, and a low-density ("gas") part, labelled $II$, both considered as homogeneous portions of nuclear matter at respective densities $\rho^I$ and $\rho^{II}$.  

The equilibrium configuration of the two phases is determined by minimizing the energy density of the cell, 
\begin{equation}
\mathcal{E} = f {\mathcal{E}}(\rho^I,\rho_p^I) + (1-f) {\mathcal{E}}(\rho^{II},\rho_p^{II}) + \mathcal{E}_{Coul} + \mathcal{E}_{surf} + \mathcal{E}_L \, ,
\label{En_dens}
\end{equation}  
which includes interface surface and Coulomb terms. 
The minimization of $\mathcal{E}$ is performed with respect to four variables: the linear size of the cluster, $R_d$, the baryonic density $\rho^I$, the proton density $\rho_p^I$ of the high-density phase, and its volume fraction $f$. 
To reduce computational time, we consider only spherical clusters, as previous studies \cite{DinhThi_2023} have shown that the inclusion of exotic geometries ("pasta" phases) has a minor effect on both the EoS and the crust-core transition point.
The Coulomb and surface terms in Eq.~\eqref{En_dens} are given by
\begin{eqnarray}
\mathcal{E}_{Coul}&=&2f e^2\pi\Phi R_d^2 \left(\rho_p^I-\rho_p^{II}\right)^2 \, , \label{eq:ecoul} \\
\mathcal{E}_{surf}&=&\frac{3f\sigma}{R_d} \label{eq:esurf}
\end{eqnarray}
where
\begin{equation}
\Phi=\frac{1}{5}\left(2-3f^{1-2/3}+f\right) \, 
\end{equation}
and the surface tension $\sigma$ reads
\begin{equation}
    \sigma=\sigma_0\frac{b+2^{4}}
    {b+y_{p,I}^{-3}+(1-y_{p,I})^{-3}} \, .
\end{equation}
Here, $y_{p,I}$ is the proton fraction of the dense phase, while the parameters $\sigma_0$ and $b$ are optimized for each model, through a fit over the measurements of the nuclear masses~\cite{Wang_2017}, as previously done in~\cite{Carreau_2019,DinhThi_2021a,DinhThi_2021b,Scurto_2024}.

\section{Analytical expressions for the NMPs} \label{App_NMP}

In this appendix we show the analytical expressions for the NMPs, including the contribution of the $\delta$ meson, used in the construction of the prior.

Due to its isovectorial nature of the $\delta$, the equation for the NMPs in the isoscalar sector remains unchanged with respect to the equations showed in \citet{Dutra_2014,Scurto_2024}, which we report here:
\begin{align}
    E_{sat}=\Bigg[ E_{kin,n}+E_{kin,p}+\frac{1}{2}m_\sigma^2\phi_0^2 
    +\frac{1}{2}m_\omega^2\omega_0^2+\frac{1}{2}m_\rho^2b_{3,0}^2\Bigg]\Bigg\vert_{n=n_{sat},n_3=0}
    \label{eq.Esat}
\end{align}
\begin{align}
    P_{sat}=0=\Bigg[P_{kin,n}+P_{kin,p}-\frac{1}{2}m_\sigma^2&\phi_0^2+\frac{1}{2}m_\omega^2\omega_0^2 \notag\\ &+\frac{1}{2}m_\rho^2b_{3,0}^2+\rho\Sigma_R\Bigg]\Bigg\vert_{n=n_{sat},n_3=0}
    \label{eq.rhosat}
\end{align}
\begin{align}
    K_{sat}=9\Bigg[n\frac{\partial\Sigma_R}{\partial n}+\frac{2g_\omega n^2}{m_\omega^2}\frac{\partial g_\omega}{\partial n}&+\frac{g_\omega^2 n}{m_\omega^2}\notag\\
    &\frac{k_F^2}{3E_F}+\frac{n M^*}{E_F}\frac{\partial M^*}{\partial n}\Bigg]\Bigg\vert_{n=n_{sat},n_3=0}
    \label{eq.Ksat}
\end{align}

In the above, the kinetic contributions to the energy per baryon are given by $E_{kin,i}=\mathcal{E}_{kin,i}/n_i$, where $\mathcal{E}_{kin,i}$ is given in the previous section.
We remind that Eq.\ref{eq.rhosat} actually connects the saturation density to the couplings by imposing the vanishing of pressure.

The equations for the NMPs in the isovector, on the other hand, are significantly modified by the presence of the new meson. We report them in the following, including also the equation for $K_{sym}$ for completion, even if it was not used for the construction of the prior.
For $J_{sym}$ we have:
\begin{align}
    J_{sym}&=\Bigg[\frac{k_F^2}{6E_F}+\frac{g_\rho^2}{8m_\rho^2}\rho-J_{sym}^\delta\Bigg]\Bigg\vert_{n=n_{sat},n_3=0}
    \label{eq.Jsym}
\end{align}
where
\begin{align}
    J_{sym}^\delta=\Bigg(\frac{g_\delta}{m_\delta}\Bigg)^2\frac{M^{*2}n}{2E_F^2(1+(g_\delta/m_\delta)^2X_\delta)},
\end{align}
\begin{align}
    X_\delta=3\Bigg(\frac{n_s}{M^*}-\frac{n}{E_F}\Bigg).
\end{align}
For $L_{sym}$ we have:
\begin{align}
    L_{sym}=\Bigg\{\frac{k_F^2}{3E_F}-\frac{k_F^4}{6E_F^3}\Bigg(1&+\frac{2M^*k_F}{\pi^2}\frac{\partial M^*}{\partial n}\Bigg)+\frac{3g_\rho^2}{8m_\rho^2}n\notag\\
    &+\frac{3g_\rho}{4m_\rho^2}n^2\frac{\partial g_\rho}{\partial n}-3nJ_{sym}^\delta A_\delta \Bigg\}\Bigg\vert_{n=n_{sat},n_3=0}
\end{align}
where
\begin{align}
    A_\delta&=\Bigg\{\frac{2}{g_\delta}\frac{\partial g_\delta}{\partial n}+B_\delta-C_\delta\Bigg\},
\end{align}
\begin{align}
    B_\delta&=\frac{M^{*2}+2M^*n(\partial M^*/\partial n)}{M^{*2}n},
\end{align}
\begin{align}
    C_\delta&=2\Bigg[2E_F\Bigg(1+\Bigg(\frac{g_\delta}{m_\delta}\Bigg)^2X_\delta\Bigg)\frac{\partial E_F}{\partial n}+E_F^2\Bigg(\frac{g_\delta}{m_\delta}\Bigg)^2\notag\\
    &\Bigg(\Bigg(\frac{2}{g_\delta}\frac{\partial g_\delta}{\partial n}+\frac{2}{M^*}\frac{\partial M^*}{\partial n}\Bigg)
    X_\delta+\frac{k_F^2}{E_F^3}\Bigg(1-\frac{3n}{M^*}\frac{\partial M^*}{\partial n}\Bigg)\Bigg)
    \Bigg]\notag\\
    &\Bigg[2E_F^2\Bigg(1+\Bigg(\frac{g_\delta}{m_\delta}\Bigg)^2X_\delta\Bigg)\Bigg]^{-1}.
\end{align}
Finally, for $K_{sym}$ we have:

\begin{align}
K_{sym}=\{9n^2(K_{sym}^\rho-K_{sym}^\delta\}\bigg\vert_{ n= n_{sat}, n_3=0}
\end{align}

where

\begin{align}
&K_{sym}^\rho=-\frac{\pi^2}{12{E_F^*}^3k_F}\left(\frac{\pi^2}{k_F} 
+ 2M^*\frac{\partial M^*}{\partial n}\right) - \frac{\pi^4}{12E_F^*k_F^4} +\frac{g_\rho}{2m_\rho^2}\frac{\partial g_\rho}{\partial n}\notag\\
&-\Bigg[\frac{\pi^4}{24{E_F^*}^3k_F^2} -
\frac{k_F\pi^2}{8{E_F^*}^5}\left(\frac{\pi^2}{k_F} +
2M^*\frac{\partial M^*}{\partial n}\right)\Bigg] \cdot 
\Bigg(1 + \frac{2M^*k_F}{\pi^2}\frac{\partial M^*}{\partial n}\Bigg) 
\notag\\
&+ \frac{ n}{4m_\rho^2}\left(\frac{\partial g _\rho}
{\partial n}\right)^2-\frac{k_F\pi^2}{12{E_F^*}^3}\cdot
\Bigg[\frac{M^*}{k_F^2}\frac{\partial
M^*}{\partial n} + \frac{2k_F}{\pi^2}\left(\frac{\partial
M^*}{\partial n}\right)^2 +\notag\\
&\frac{2k_F M^*}{\pi^2}\frac{\partial^2
M^*}{\partial n^2}\Bigg]
+\frac{g_\rho n}{4m_\rho^2}\frac{\partial^2 g_\rho}{\partial n^2}
\label{eq.Ksym}
\end{align}

and

\begin{align}
K_{sym}^\delta=J_{sym}^\delta\Bigg\{A_\delta^2+2\Bigg[\frac{1}{g_\delta}\frac{\partial^2g_\delta}{\partial n^2}-\Bigg(\frac{1}{g_\delta}\frac{\partial g_\delta}{\partial n}\Bigg)^2\Bigg]+2D_\delta-B_\delta^2+C_\delta^2
-E_\delta
\Bigg\},
\end{align}

with
\begin{align}
    D_\delta&=\frac{2M^{*}(\partial M^*/\partial n)+n(\partial M^*/\partial n)^2+M^*n(\partial^2 M^*/\partial n^2)}{M^{*2}n},
\end{align}

\begin{align}
    &E_\delta=2\Bigg[2\Bigg(\frac{\partial E_F}{\partial n}\Bigg)^2\Bigg(1
    +\Bigg(\frac{g_\delta}{m_\delta}\Bigg)^2X_\delta\Bigg)+2E_F\Bigg(1
    +\Bigg(\frac{g_\delta}{m_\delta}\Bigg)^2X_\delta\Bigg)\frac{\partial^2 E_F}{\partial n^2}\notag\\    
    &+4E_F\frac{\partial E_F}{\partial n}\Bigg(\frac{g_\delta}{m_\delta}\Bigg)^2
    \Bigg(\Bigg(\frac{2}{g_\delta}\frac{\partial g_\delta}{\partial n}+\frac{2}{M^*}\frac{\partial M^*}{\partial n}\Bigg)
    X_\delta+\frac{k_F^2}{E_F^3}\Bigg(1-\frac{3n}{M^*}\frac{\partial M^*}{\partial n}\Bigg)\Bigg)\notag\\
    &+2E_F^2\Bigg(\frac{g_\delta}{m_\delta}\Bigg)^2F_\delta\Bigg]\Bigg[2E_F^2\Bigg(1+\Bigg(\frac{g_\delta}{m_\delta}\Bigg)^2X_\delta\Bigg]^{-1},
\end{align}

\begin{align}
    &F_\delta=X_\delta\Bigg(\frac{1}{g_\delta}\frac{\partial^2g_\delta}{\partial n^2}+\Bigg(\frac{1}{g_\delta}\frac{\partial g_\delta}{\partial n}\Bigg)^2\Bigg)+2\frac{1}{g_\delta}\frac{\partial g_\delta}{\partial n}\Bigg(X_\delta\frac{2}{M^*}\frac{\partial M^*}{\partial n}\notag\\
    &+\frac{k_F^2}{E_F^3}\Bigg(1-\frac{3n}{M^*}\frac{\partial M^*}{\partial n}\Bigg)\Bigg)+\frac{1}{2}\Bigg[2X_\delta\Bigg(\frac{1}{M^*}\frac{\partial^2 M^*}{\partial n^2}-\frac{1}{M^{*2}}\Bigg(\frac{\partial M^*}{\partial n}\Bigg)\Bigg)\notag\\
    &+\frac{2}{M^*}\frac{\partial M^*}{\partial n}\Bigg(X_\delta\frac{2}{M^*}\frac{\partial M^*}{\partial n}
    +\frac{k_F^2}{E_F^3}\Bigg(1-\frac{3n}{M^*}\frac{\partial M^*}{\partial n}\Bigg)\Bigg)\notag\\
    &-\frac{3k_F^2}{E_F^3}\Bigg(\frac{1}{M^*}\frac{\partial M^*}{\partial n}+\frac{n}{M^*}\frac{\partial^2 M^*}{\partial n^2}-\frac{n}{M^{*2}}\Bigg(\frac{\partial M^*}{\partial n}\Bigg)^2\Bigg)+\Bigg(1-\frac{3n}{M^*}\frac{\partial M^*}{\partial n}\Bigg)\notag\\
    &\Bigg(\frac{2k_F}{E_F^3}\frac{\partial k_F}{\partial n}-\frac{3k_F^2}{E_F^4}\frac{\partial E_F}{\partial n}\Bigg)\Bigg]
\end{align}

\end{appendix}

\end{document}